Imaging the Effects of Individual Zinc Impurity Atoms

on Superconductivity in $Bi_2Sr_2CaCu_2O_{8+\delta}$

S.H. Pan[1], E.W. Hudson[1], K.M. Lang[1], H. Eisaki[2], S. Uchida[2], and J.C. Davis[1]

1.  Department of Physics, University of California, Berkeley, CA 94720, USA.

2.  Department of Superconductivity, University of Tokyo, Tokyo, Japan.

**Although their crystal structures are complex, all high temperature superconductors contain some crystal planes consisting of only Cu and O atoms in a square lattice. Superconductivity is believed to originate from strongly interacting electrons in these $CuO_2$ planes. Substitution of a single impurity atom at a Cu site creates a simple but powerful perturbation to these interactions. Detailed knowledge of the effects of such an impurity atom on the superconducting order parameter and on the quasi-particle local density of states (LDOS) could allow competing theories of high temperature superconductivity (HTSC) to be tested at the atomic scale. The fundamental implications of results from numerous bulk measurements on samples doped with impurity atoms could also be clarified with such data. Here we describe scanning tunneling microscopy studies of the effects of individual Zn impurity atoms located at the Cu site in the high-$T_c$ superconductor $Bi_2Sr_2CaCu_2O_{8+d}$ Tunneling spectroscopy shows intense quasi-particle scattering resonances[1] at the Zn sites, coincident with strong suppression of superconductivity within about 15 Å. Imaging of the quasi-particle LDOS at these sites reveals the long sought four-fold symmetric "quasi-particle cloud" aligned with the d-wave gap**

**nodes. Several unexpected phenomena, which can shed new light on the atomic-scale response of HTSC to a probe impurity atom, are also observed.**

A single impurity atom, substituted for Cu in the $CuO_2$ plane, is a perturbation that strongly disrupts the surrounding electronic environment at the atomic scale. Since it can therefore be an ideal tool to probe the mechanism and phenomenology of high temperature superconductivity, numerous theoretical studies[2-11] have predicted the local effects of an impurity atom on the superconducting order parameter and on the spatial dependence of the quasi-particle LDOS. However, until now, no experimental data has been available for direct comparison with these predictions.

Zn doped high temperature superconductors exhibit strongly altered bulk superconducting properties, including reductions of the critical temperature[12-17] and of the superfluid density[13]. Explanations for these phenomena, as well as for increased in-plane resistivity[12] and density of states at the Fermi level[14], and reduced microwave surface resistance[15], have been based on proposals of very strong quasi-particle scattering rates and dramatic reduction in the order parameter near the Zn site. Spectroscopic features in the infrared conductance have been interpreted in terms of localized quasi-particle states, with non-zero energy, at the Zn sites[17]. In addition, with Zn impurity atoms present, ARPES experiments show dramatic changes in the quasi-particle features for some **k**-space directions[18], and neutron scattering unexpectedly shows persistence of the strong antiferromagnetic correlations far above $T_c$[19]. The implications of these results for the fundamentals of HTSC could be greatly clarified with data on the atomic-scale effects of individual Zn atoms on high temperature superconductivity.

In this paper we describe the first scanning tunneling microscopy (STM) experiments on the effects of individual Zn impurity atoms in $Bi_2Sr_2Ca(Cu_{1-x}Zn_x)_2O_{8+\delta}$ (Zn-BSCCO). Previous experiments have shown STM to be an excellent tool for study of the effects of individual impurity atoms in several other systems[20-24]. For our studies, we use a custom-built, low temperature STM[25] which can simultaneously image both the surface topography and the electronic LDOS.

The Zn-BSCCO single crystals used in these experiments are grown by the floating zone method[26] with an x = 0.6% nominal doping concentration of Zn. They have been characterized to have a $T_c$ of 84 K with a transition width of 4 K. These crystals are cleaved in-situ at 4.2 K in cryogenic ultrahigh vacuum and are inserted into the STM.

Immediately after cleavage, a topographic image is taken to determine the condition of the crystal surface, and a typical result is shown in Fig. 1a. A single crystal layer, believed to usually be the BiO plane with only the Bi atoms apparent in STM imaging[27], is exposed. The indistinguishability of these surfaces from those of non Zn-doped BSCCO[1,27-28] is consistent with the Zn dopant atoms being at the Cu sites (two layers below the exposed surface).

To search for low energy quasi-particle states associated with the Zn atoms, we next map the differential conductance of the surface at zero sample bias. Figure 1b is a typical resulting LDOS image taken on a 500 Å square region. It shows an overall dark background indicative of a very low quasi-particle LDOS near the Fermi level. This is as expected for a superconductor far below $T_c$. Remarkably however, a number of randomly distributed bright sites with high LDOS and a distinct four-fold symmetry are observed.

Further investigation of these bright sites reveals striking differences between conductance spectra taken exactly at their centers and spectra taken at regular superconducting regions of the sample (dark areas in Fig. 1b). In Fig. 2 we show two such spectra, from which several significant observations can be made. First, the spectrum at the center of a typical bright site has an LDOS peak which is up to six times greater than the normal-state conductance. Second, the peak occurs at an energy $\Omega$ low compared to the gap energy, with typical values $\Omega = -1.5 \pm 0.5$ mV. Third, at these sites the superconducting coherence peaks (identified by the arrows in Fig. 2) are strongly suppressed, indicating the almost complete destruction of superconductivity. All of these phenomena are among the theoretically predicted characteristics of very strong quasi-particle scattering in a d-wave superconductor[4,5,7,8,10].

To identify the source of these resonances, simultaneous pairs of high resolution topographic and LDOS images, centered on individual resonance sites, are next acquired. Comparison between pairs of these images shows that the center of a scattering resonance always coincides with the site of a surface Bi atom, as shown for example in Figures 3a and 3b. From the known crystal structure of BSCCO, a Cu site, where the Zn dopant atom could reside, is directly below each exposed Bi atom, and separated from it by the intervening $SrO_2$ layer. Furthermore, the density of the observed scattering resonances is $x = 0.2 \pm 0.1$ %, in reasonable agreement with the nominal Zn doping concentration[29]. Finally, LDOS features like those in Fig. 3b have never been observed in non-Zn-doped BSCCO crystals. We therefore attribute these strong resonances to quasi-particle scattering at the Zn impurity atoms.

High resolution LDOS images of scattering sites at the resonance energy show structures with very strong central peaks whose magnitude can be more than 100 times higher than the background. Although these LDOS structures are highly localized, nonetheless they have variations in intensity over two orders of magnitude. To demonstrate these elements in full detail, Fig. 3b is presented in a logarithmic intensity scale. In this representation we see that, in addition to the central peak, a dominant feature is a relatively bright cross-shaped region aligned along the crystal **a**- and **b**-axes, and extending to a radius of about 10 Å. Concentric with this bright cross, and rotated 45 degrees relative to it, another cross-shaped feature with considerably lower intensity extends to approximately 30 Å from the center.

The LDOS does not decay monotonically with distance $r$ from the scatterer but rather oscillates, producing local minima and maxima. To help clarify the register of these LDOS oscillations to the crystal, we show schematically in Fig. 3c the positions of the Cu and O atoms in the $CuO_2$ plane with the relative positions of the LDOS maxima overlaid as heavy circles. Also shown are the orientations of the crystal **a**- and **b**- axes, and that of the d-wave gap nodes as determined by ARPES[30]. Analysis of the simultaneously imaged atomic positions and LDOS intensities in Fig. 3a and 3b indicates that the LDOS maxima coincide with positions of only some of the atomic sites near the scattering center. The positions of the four nearest-neighbor Cu atoms to the Zn have no local LDOS maxima associated with them. On the other hand, the positions of the eight second-nearest-neighbor and third-nearest-neighbor Cu atoms coincide with local maxima in the LDOS and appear to form a 'box' around the scatterer as seen in Fig. 3b.

In Fig. 4a the normalized LDOS at the resonance energy is plotted as a function of $r$ along both the gap node and maxima directions. Due to the strength of the oscillations it is difficult to identify a power law governing the falloff of LDOS with $r$ in these directions. However, the rate of decay is clearly faster in the gap node directions, while towards the gap maxima weak "quasi-particle beams" are apparent up to 30 Å from the scatterer.

The Zn atom affects not only the low energy LDOS, but also features at the gap energy scale. Its effect on the magnitude of the coherence peaks, and on the width of the superconducting gap, is shown in Fig. 4b which is a plot of the measured evolution of the complete LDOS spectrum as the tip is moved away from the scattering site along the **b**-axis. The superconducting coherence peaks are strongly suppressed at the Zn scattering site and recover to their bulk value over a distance of about 15 Å.

We observe all of the above phenomena to be characteristic of large numbers of scattering sites in several Zn-doped BSCCO crystal samples. Some of these phenomena are in very good agreement with theoretical predictions. For example, theoretical proposals that quasi-particle scattering resonances, and localized impurity states, could be created by *individual* impurity atoms in a *d*-wave superconductor[3-5,7-11] are here shown to be correct. The strength of the scattering leading to these resonances, in terms of a phase shift $\delta_0$, may be calculated, in the context of Refs. 4 and 5, using the ratio of the resonance energy $\Omega$ to the energy gap $\Delta_0$:

$$\left|\frac{\Omega}{\Delta_0}\right| \approx \frac{\pi/2}{\ln(8/\pi)} \qquad (1)$$

where c = cot($\delta_0$). Using an average gap value (away from the impurities) of $\Delta_0$ = 44 mV and $\Omega$ = -1.5 mV, we obtain a phase shift of 0.48$\pi$. This confirms that scattering from Zn is very close to the unitary limit ($\delta_0 = \pi/2$)[2-11,15-16]. The predicted particle-hole symmetry breaking for an impurity atom scattering resonance[4,5] is directly confirmed, while the predicted four-fold symmetry in the LDOS, and its alignment with the d-wave gap nodes, near an impurity atom[3,4,5] are also clearly observed.

These measurements also provide the first microscopic validation for models proposed to explain the results of several bulk experiments. A proposal that superfluid density reductions can be explained by non-superconducting regions of area $\pi\xi^2$ around each impurity atom[13] (the "Swiss cheese" model) is directly validated by the suppression of superconductivity near the Zn site shown in Fig 4b. The strong *non-zero energy* quasi-particle LDOS at the impurity atoms, which was postulated to explain the infrared results[17], is also directly confirmed by the results in Fig. 2.

Despite this consistency between previous theoretical and experimental results and our measurements, several unexpected phenomena are observed. These include the register of the "quasi-particle cloud" to the atomic sites, and the orientation of "quasi-particle beams" with the gap maxima. These phenomena may indicate that effects such as band structure[9,11], local magnetic moments at the Zn site[31,32], and tunneling through intermediate atomic layers, need to be considered. They may, however, also be relevant to novel theories of HTSC. Although the oscillations in the LDOS were expected at the Fermi wavelength[3-5,7,8,11], correspondence is instead found between locations of some specific atoms, and local maxima of the LDOS oscillations. These correspondences might indicate that the strong correlation effects, upon which the HTSC is probably based, are being observed directly. For example, in Zn-doped HTSC samples neutron scattering shows antiferromagnetic correlations (which usually only exist below $T_c$) persisting well above $T_c$[19]. This might be explained if the Zn atom creates antiferromagnetic order locally. The LDOS oscillations observed here by STM have wavevector ($\pi/a$, $\pi/a$), which is identical to the wavevector of antiferromagnetic order seen in the neutron scattering results[19]. Thus, our results may provide a new probe of the relationship between superconductivity and antiferromagnetic order, and therefore be relevant to the SO(5) theory of HTSC at the atomic-scale[33].

In conclusion, we report the first STM studies of the effects on high temperature superconductivity of individual impurity atoms substituted at the Cu site in the $CuO_2$ plane of BSCCO. Associated with the Zn impurities we find intense quasi-particle resonances consistent with unitary scattering in a d-wave superconductor. LDOS imaging at the resonance energy shows a highly localized "quasi-particle cloud" which has a clear four-fold symmetry aligned with the d-wave gap nodes, in qualitative agreement with theory. We report other observations that directly validate the deductions of a variety of previous theoretical and experimental studies. New phenomena are also reported which are not described by existing theories and can thus provide new information on the physics of HTSC at the atomic-scale. Finally, the demonstration of quasi-particle scattering spectroscopy and LDOS imaging at individual impurity atoms in the cuprate-oxides opens a new avenue for research into these important materials.

We acknowledge and thank A. Balatsky, M. Crommie, M. Flatté, M. Franz, S. Kashiwaya, A. de Lozanne, A. MacDonald, V. Madhavan, M. Ogata, J. Orenstein, D. J.

Scalapino, Z.-X. Shen, and Y. Tanaka, for helpful conversations and communications. This work was supported by the LDRD Program of the Lawrence Berkeley National Laboratory under the Department of Energy Contract No. DE-AC03-76SF00098, by the D. & L. Packard Foundation, by Grant-in-Aid for Scientific Research on Priority Area (Japan), and by a COE Grant from the Ministry of Education, Japan. Correspondence and requests for materials should be addressed to J.C. Davis, jcdavis@socrates.berkeley.edu.

Figure **1a.**     A 150 Å square constant current topograph of a Zn-BSCCO single crystal. The atoms are displaced from their ideal square lattice sites, forming a supermodulation along the crystal **b**-axis. This surface is indistinguishable from typical BiO cleavage-plane surfaces of non Zn doped BSCCO. (T = 4.2 K, I = 100 pA, $V_{sample}$ = -100 mV).

Figure **1b**     A 500 Å square differential tunneling conductance map taken at sample bias $V_{sample}$ = 0 on the same surface (with the same orientation) as that shown in Fig. 1a. Since the differential tunneling conductance dI/dV is proportional to the sample LDOS this is also a map of the quasi-particle LDOS at the Fermi energy ($V_{sample}$ = 0). Most of the image appears dark, which is indicative of a very low density of quasi-particles at the Fermi level. The Zn scattering sites appear as bright regions of approximate dimension 15 Å, each clearly exhibiting a cross-shaped four-fold symmetric structure. Note that the orientation of all these cross-shaped features is the same. To acquire this image we operate at 4.2 K, set a 1 GΩ junction resistance (I = 200 pA, $V_{sample}$ = -200 mV), and measure the conductance with a standard low frequency ac lock-in technique ($A_{modulation}$ = 500 $\mu V_{rms}$, $f_{modulation}$ = 447.3 Hz).

Figure **2**.     Differential tunneling conductance versus sample bias taken at two different locations on the Zn-BSCCO crystal. The spectrum of a 'regular' superconducting region of the sample, where Zn scatterers are absent (dark in Fig 1b), is shown as the solid circles. The superconducting coherence peaks are indicated by the arrows. The data shown as open circles, with an interpolating fine solid line, is the spectrum taken exactly at the center of a bright scattering site. It shows both the intense scattering resonance centered at $\Omega = -1.5$ mV, and the very strong suppression of the superconducting coherence peaks and gap magnitude at this same location.

Figure **3a** and **3b**

Figures 3a and 3b are simultaneously acquired, 60 Å square, high spatial resolution images of the topography and the differential conductance at $V_{sample} = -1.5$ mV. LDOS imaging at other energies from $-3$ mV to $3$ mV shows qualitatively similar phenomena. The bright center of the scattering resonance as seen in Fig. 3b is located at the position of the Bi atom marked by an X in Fig. 3a. Note that Fig. 3b uses a logarithmic scale of intensity since the features at large distance from the Zn site are two orders of magnitude weaker than the resonance peak. The inner bright cross is oriented with the nodes of the d-wave gap (as indicated schematically in Fig. 3c). The weaker outer features, including the ~30 Å long "quasi-particle beams" at 45 degrees to the inner cross, are oriented with the gap maxima.

Figure **3c**   This is a 30 Å square schematic representation of the square $CuO_2$ lattice, showing its relative orientation to the BiO surface in Fig. 3a. The Cu atoms are indicated as large solid circles and the O atoms as small solid circles, while the Zn scatterer is at the central point. The crystal **a**-axis and **b**-axis orientations, and the orientations of the maxima and nodes of the d-wave superconducting gap, as measured by ARPES experiments, are also shown. The location of the maxima in the LDOS, as measured from Fig. 3b, are shown schematically as heavy circles surrounding the atomic sites with which they appear to be associated. The intensity of the LDOS at these sites is shown schematically by the thickness of the line forming these circles.

Figure **4a**   The differential tunneling conductance (normalized to the peak value) versus distance $r$ from the center of a Zn scatterer, along the **a**-axis or **b**-axis (gap-node) directions (shown as open circles) and along the Cu-O bond (gap-maxima) directions (shown as open squares). Each data point is the LDOS at distance $r$ averaged over an inclusive angle of $10^o$ around the given directions. The data are typical of averaged high-resolution conductance maps at isolated Zn scatterers.

Figure **4b**   A series of tunneling spectra taken on a line from the scatterer along the **b**-axis direction in 0.5 Å steps. This clearly shows suppression of the coherence peaks (whose absence at the Zn scatterer is indicated by upward pointing arrows), and recovery of the superconductivity (along with the coherence peaks whose presence is indicated by downward pointing arrows) within a distance of ~ 15 Å from the scattering site.

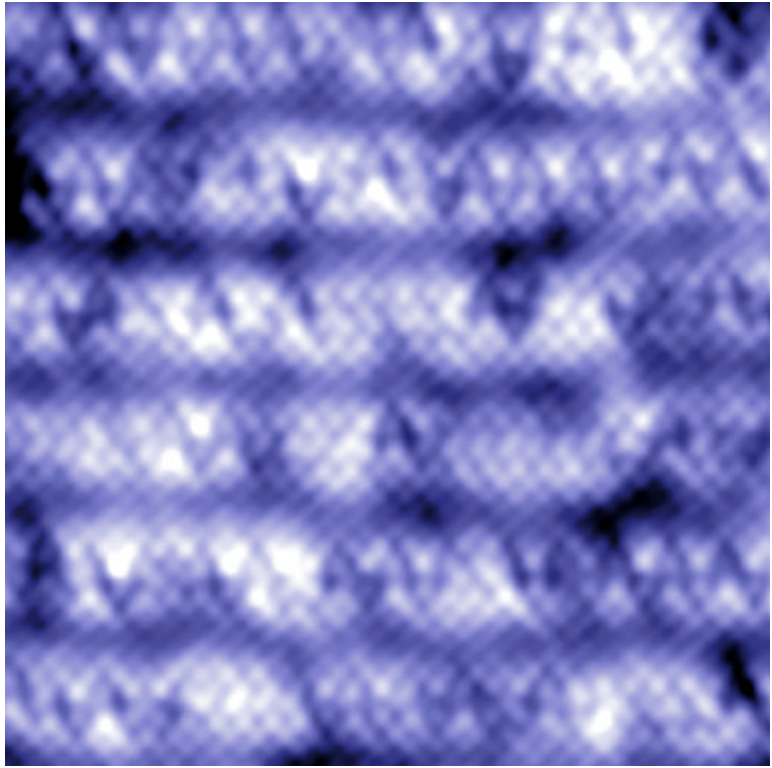
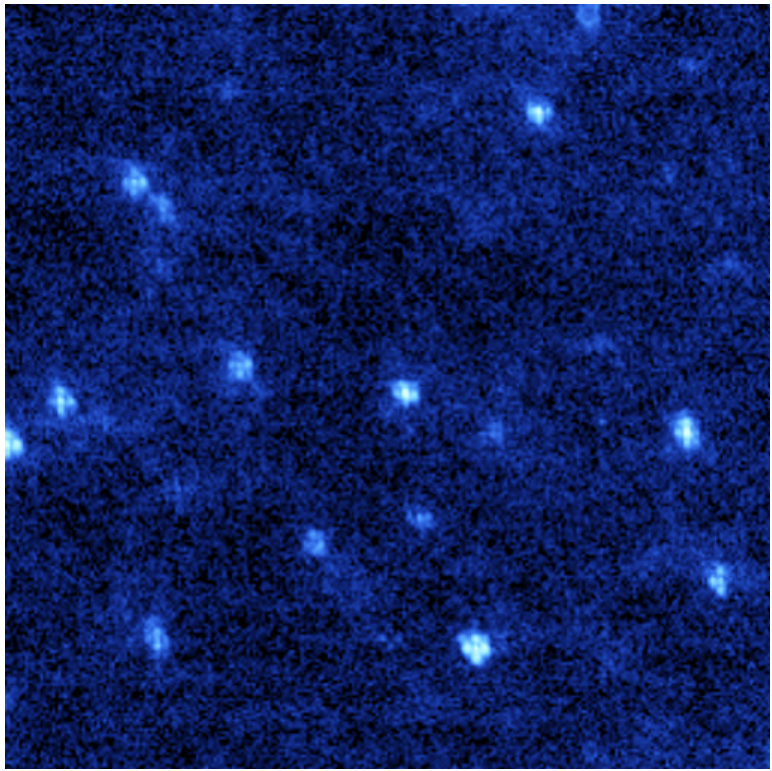

Figure 1
J.C. Davis

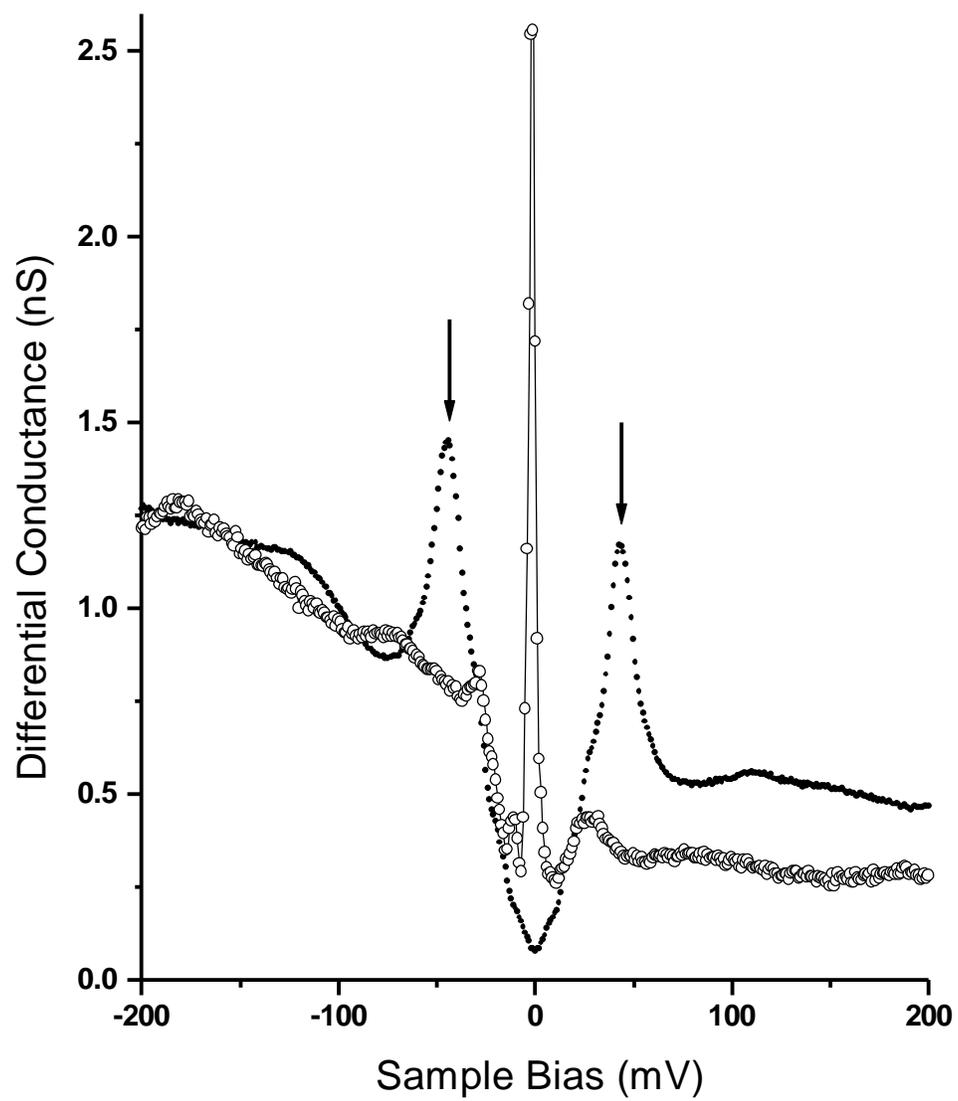

Figure 2
J.C. Davis

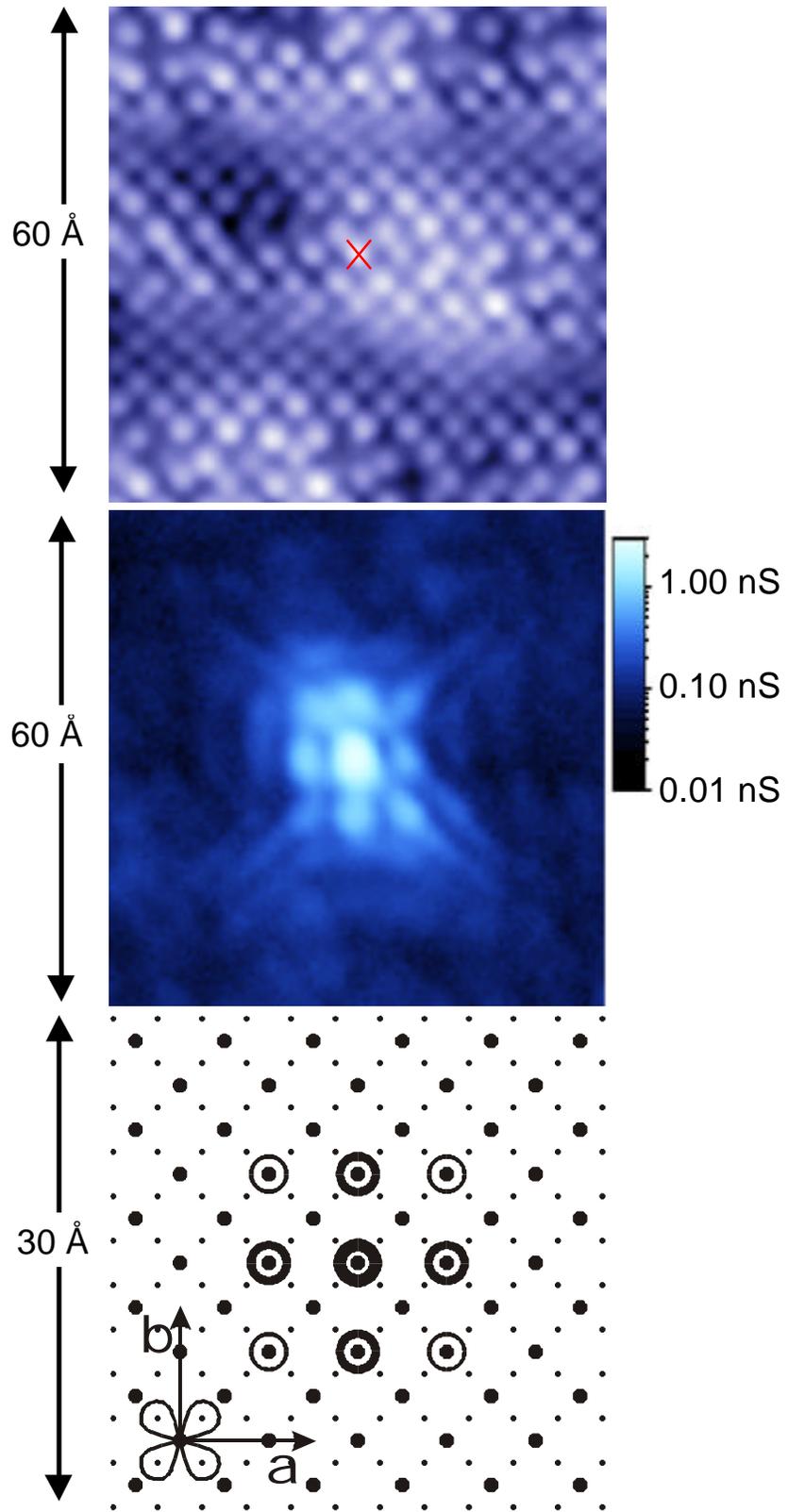

Figure 3
J.C. Davis

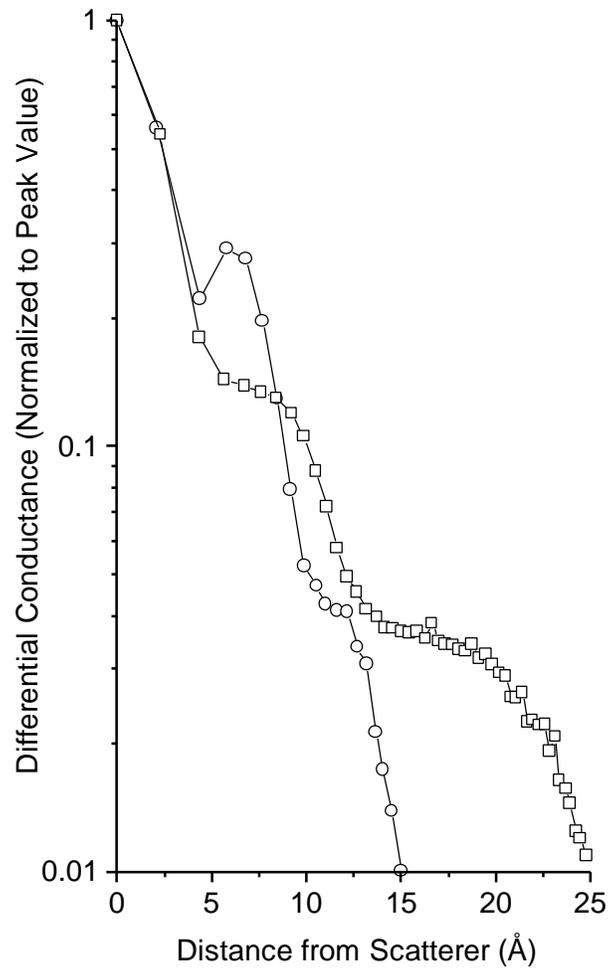

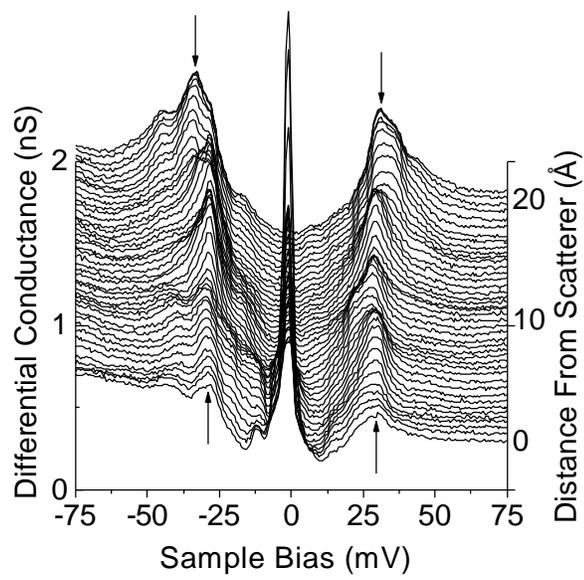

Figure 4
J.C. Davis